\documentclass[3p,times,procedia,secnumarabic, graphics,floatfix, nofootinbib,tightenlines,nobibnotes, aps, prl, 12pt]{elsarticle}
\usepackage{graphicx}
\usepackage{amsmath,amssymb}
\usepackage{amsfonts}
\usepackage{multirow}

\biboptions{authoryear}

\begin{document}

\title{Dynamical capacity drop in a nonlinear stochastic traffic model}

\author{Wei-Liang Qian$^{a,b}$, Adriano F. Siqueira$^{a}$, Romuel  F. Machado$^{c}$, Kai Lin$^{a}$ and Ted W. Grant$^{a}$}

\address[a]{Departamento de Ci\^encias B\'asicas e Ambientais, Escola de Engenharia de Lorena, Universidade de S\~ao Paulo, 12602-810, Lorena, SP, Brazil}
\address[b]{Departamento de F\'isica e Qu\'imica, Faculdade de Engenharia de Guaratinguet\'a, Universidade Estadual Paulista, 12516-410, Guaratinguet\'a, SP, Brazil}
\address[c]{Departamento de F\'{i}sica, Universidade Federal de Ouro Preto, 45300-000, Ouro Preto, MG, Brazil}

\date{Sept. 20, 2017}

\begin{abstract}
In this work, we show that the inverse-$\lambda$ shape in the fundamental diagram of traffic flow can be produced dynamically by a simple nonlinear mesoscopic model with stochastic noises.
The proposed model is based on the gas-kinetic theory of the traffic system.
In our approach, the nonlinearity leads to the coexistence of different traffic states.
The scattering of the data is thus attributed to the noise terms introduced in the stochastic differential equations and the transition among the various traffic states.
Most importantly, the observed inverse-$\lambda$ shape and the associated sudden jump of physical quantities arise due to the effect of stochastic noises on the stability of the system.
The model parameters are calibrated, and a qualitative agreement is obtained between the data and the numerical simulations.
\end{abstract}

%\pacs{PACS numbers: 89.40.-a, 47.85.Dh, 05.60.Cd, 05.40.-a}
\maketitle

\section{Introduction}

Based on the empirical observations accumulated on highways in different countries for nearly 8 decades (see for instance the Refs.\citep{traffic-flow-review-02,traffic-flow-data-01,traffic-flow-phenomenology-01,traffic-flow-data-inverse-lambda-05}), one important empirical feature of the so called {\it fundamental diagram} of traffic flow is the inverse-$\lambda$ shape \citep{traffic-flow-data-inverse-lambda-01,traffic-flow-data-inverse-lambda-02,traffic-flow-data-inverse-lambda-03,traffic-flow-data-inverse-lambda-04} accompanying with capacity drop \citep{traffic-flow-data-capacity-drop-01,traffic-flow-data-capacity-drop-02,traffic-flow-data-capacity-drop-03}.
The inverse-$\lambda$ implies discontinuity of the flow as a function of vehicle density which occurs in the vicinity of the maximum of the flow.
Consequently, the flow-concentration curve is divided into two different regions of lower and higher vehicle density respectively, known as {\it free} and {\it congested} flow.
However, further data analyses \citep{traffic-flow-data-inverse-lambda-05,traffic-flow-data-inverse-lambda-06,traffic-flow-data-inverse-lambda-07} give birth to an intriguingly different viewpoint.
When one divides the empirically observed data into stationary and non-stationary traffic conditions, the resulting fundamental diagram is continuous and scatter-free when one is only interested in the {\it average values} of sustained periods of nearly {\it stationary} traffic conditions.
In \citep{traffic-flow-data-inverse-lambda-05}, each point in the scatterplot in question corresponds to the average properties of prolonged periods ($\sim 10$ min) of near-stationary traffic states.
It exhibits a well-defined relation between flow and concentration, which does not show evidence of discontinuities.
The resulting scatterplot from the near-stationary data is understood to be different from that obtained by simply collecting consecutive time intervals of fixed duration.
This is because the latter includes the information of non-stationary states.
%While the discontinuity measures the difference between free and congested traffic states, the observed scatter is associated with the non-stationary transitions between the two traffic states.
Therefore, a part of the observed complexity in the fundamental diagram could be understood as a result of mixing traffic data from stationary and non-stationary traffic conditions. 
The above analysis and the associated reasoning procedure are well recognized and adopted by many authors (eg. \citep{traffic-flow-micro-22,traffic-flow-phenomenology-13,traffic-flow-hydrodynamics-19,traffic-flow-cellular-11}).
Moreover, further developments including new methods and models were proposed, and debate continues on how to address the scatter \citep{traffic-flow-data-17,traffic-flow-data-18}, and the capacity drop \citep{traffic-flow-hydrodynamics-19,traffic-flow-hydrodynamics-20}.

As a matter of fact, modeling of the scatter in the fundamental diagram has always attracted much attention in the study of traffic flow.
From an empirical viewpoint, the uncertainties observed in the data can always be expressed in terms of the variance of the fundamental diagram, which has become an attractive topic in recent years \citep{traffic-flow-hydrodynamics-10,traffic-flow-micro-12,traffic-flow-hydrodynamics-11,traffic-flow-review-05,traffic-flow-btz-05}.
To study such uncertainties, methodologies involving stochastic modeling have aroused many discussions, either from macroscopic viewpoint \citep{traffic-flow-phenomenology-02,traffic-flow-phenomenology-03,traffic-flow-phenomenology-04,traffic-flow-phenomenology-05}, from microscopic models \citep{traffic-flow-micro-13,traffic-flow-micro-14,traffic-flow-micro-15,traffic-flow-cellular-01,traffic-flow-cellular-05,traffic-flow-micro-16,traffic-flow-cellular-07,traffic-flow-cellular-08,traffic-flow-cellular-09,traffic-flow-cellular-10} or from phenomenological approaches \citep{traffic-flow-phenomenology-06,traffic-flow-phenomenology-07,traffic-flow-phenomenology-08}.
On the one hand, even under stationary traffic conditions, a certain degree of residual scatter always persists. 
They are, for the most part, statistical in nature.
The cause of such uncertainty can be attributed to the heterogeneous drivers \citep{traffic-flow-data-10} and the lane changes \citep{traffic-flow-phenomenology-12} among others.
On the other hand, when the fundamental diagram is obtained by using a smaller aggregation interval, one may extract information on the intrinsically non-stationary state or the dynamics of transitions between different traffic states.
These non-stationary states can be either deterministic or stochastic.
In literature, many parsimonious models were developed in their efforts to understand the underlying dynamics. 
Some of these traffic conditions are understood to be connected to the instability of the equation of motion (EoM) \citep{traffic-flow-hydrodynamics-04,traffic-flow-micro-07,traffic-flow-micro-08,traffic-flow-micro-09,traffic-flow-micro-17} and subsequent phase transitions in the system \citep{traffic-flow-catastrophe-01,traffic-flow-catastrophe-02,traffic-flow-review-01,traffic-flow-cellular-02, traffic-flow-cellular-10}.
The scatter of the data can be either attributed to the (transient) temporal evolution of the dynamical system (such as stop-and-go wave \citep{traffic-flow-hydrodynamics-04,traffic-flow-hydrodynamics-07,traffic-flow-hydrodynamics-15,traffic-flow-hydrodynamics-13,traffic-flow-hydrodynamics-14,traffic-flow-hydrodynamics-12,traffic-flow-data-11}), or to a metastable region in the fundamental diagram (such as the {\it synchronized flow} defined in the three phase traffic theory \citep{traffic-flow-three-phase-01,traffic-flow-three-phase-02}).

The present work involves an attempt to show that both the inverse-$\lambda$ shape and its disappearance can be understood within a nonlinear mesoscopic model with stochastic noise.
We argue that underlying physical mechanism is the modified stability of the system owing to the stochastic noises.
A mesoscopic model \citep{traffic-flow-btz-01,traffic-flow-btz-02,traffic-flow-btz-03,traffic-flow-btz-04} seeks a compromise between the microscopic and the macroscopic approaches.
Instead of focusing on individual vehicles, the model describes traffic flow in terms of vehicle distribution as a function of space–time coordinate and vehicle speed. 
The dynamics of the system is therefore determined by the EoM for the distribution function, which usually takes the form of an integro-differential equation such as the Boltzmann equation.
In fact, most mesoscopic models can be derived in analogy to the gas-kinetic theory.
As it will be discussed below, for near-stationary traffic states, observable traffic states correspond to stable solutions of the deterministic EoM, the resultant fundamental diagram is continuous, and the deviation is small.
For non-stationary traffic states, on the other hand, random noises modify stability of the corresponding stochastic EoM, and the resulting fundamental diagram becomes discontinuous and shows an inverse-$\lambda$ shape.
Our mesoscopic approach considers discrete speed states. 
It makes use of the stochastic differential equations (SDE) to model the system with random noises.

The paper is organized as follows.
In the next Section, we present a simplified deterministic mesoscopic model characterized by a {\it fold catastrophe} potential function.
It is shown that the model leads to two different traffic phases, and the two phases join continuously at the maximum of flow.
In Section 3, the model is generalized to include stochastic elements. 
We investigate the stability of the solutions of the SDE numerically and study the variance of the flow. 
It turns out that the appearance of the discontinuity in the fundamental diagram is obtained naturally without introducing any additional parameter.
In Section 4, the present model is then compared qualitatively to the data from Interstate 80 (I-80) freeway collected under the NGSIM program.
Concluding remarks and perspectives are given in Section 5 of the paper.

\section{A deterministic fold catastrophe model}

Recently, we proposed a stochastic mesoscopic traffic model \citep{traffic-flow-btz-lob-01} and employed it to study the fundamental diagram and the flow variance. 
The model is based on the gas-kinetic approach for the traffic system, initially introduced by Prigogine and Andrews \citep{traffic-flow-btz-01}.
In our study, we consider a small section of highway where homogeneity can be assumed.
As an approximation, the vehicle speed is discretized, similar to the role of space in the automaton model \citep{traffic-flow-cellular-01}.
Therefore, the degree of freedom of the system is the vehicle accumulation $n_i$ of the $i$-th speed state.
According to the mesoscopic description of the traffic system, the EoM determines the transition rates between different speed states.
The mathematical simplicity of the model provides the facility to acquire all the solutions of the EoM analytical, and subsequently, to carry out the analysis of the stability of the system as well as its relationship to traffic congestion (c.f. \citep{traffic-flow-btz-lob-02}).

In \citep{traffic-flow-btz-lob-01}, only the linear transition terms were considered, and the stability of the system was not studied.
In this section, the above model is generalized to include the lowest order nonlinear contributions, and all the stationary solutions are obtained. 
We note that a stationary solution is not necessarily a stable one.
In the context of a traffic system, only stable solutions and quasi-stationary solutions are relevant and can be associated with observable traffic states.
As discussed below, the two stationary solutions of the model are found to be conditionally stable and turn out to correspond to the two fundamental traffic states, namely, the free flow and the congested flow states.
Although it is built in terms of a minimal number of parameters, we argue that the proposed model reproduces the essential characteristics of the fundamental diagram.
We will also show this model is equivalent to a fold catastrophe model.

For a section of highway of length $L$ which contains $N$ vehicles, the EoM of the present model is essentially a simplified transport equation, given as follows \footnote{A preliminary version of the model can be found in \citep{traffic-flow-btz-lob-03}.}
\begin{eqnarray}
\frac{dn_1}{dt}&=&-c_1n_1+c_2n_1 n_2\frac{1}{N_{max}-N} ,\nonumber \\
\frac{dn_2}{dt}&=& -c_2n_2 n_1\frac{1}{N_{max}-N}+c_1n_1 ,
\label{eom1}
\end{eqnarray}
where $n_i$ ($i=1,2$) is the vehicle accumulation of the $i$-th state with speed $v_i$ and $N_{max}$ is maximum accumulation owing to finite vehicle size.
So $k_{max}=N_{max}/L$ is the corresponding maximal vehicle density leading to complete congestion.
Without loss of generality, we assume $v_1 < v_2$.
 
The physical contents of the EoM and the coefficients are the following.
The transition rate for the speed state $1$ is determined by the r.h.s. of the equation for $n_1$ which consists of a loss term and a gaining term.
The loss term $c_1n_1$ is simply proportional to the vehicle accumulation since cautious slow drivers leave this state at a fixed rate.
In addition to being proportional to $n_2$, the gain term must also be proportional to $n_1$, since slow drivers force the fast ones to slow down to their speed.
The extra constant factor $\frac{1}{N_{max}-N}$ takes the total congestion into account as will be explained below.
It is easy to show that Eq.(\ref{eom1}) guarantees the total vehicle number conservation $N=n_1+n_2$ and hence there is only one degree of freedom.
Therefore one may simply consider the following equation in terms of $n_1$.
\begin{eqnarray}
\frac{dn_1}{dt}\equiv f(n_1)=-c_1n_1+c_2n_1 \frac{N-n_1}{N_{max}-N} .
\end{eqnarray}

\begin{figure}[!htb]
\begin{tabular}{cc}
\begin{minipage}{200pt}
\centerline{\includegraphics*[width=8cm]{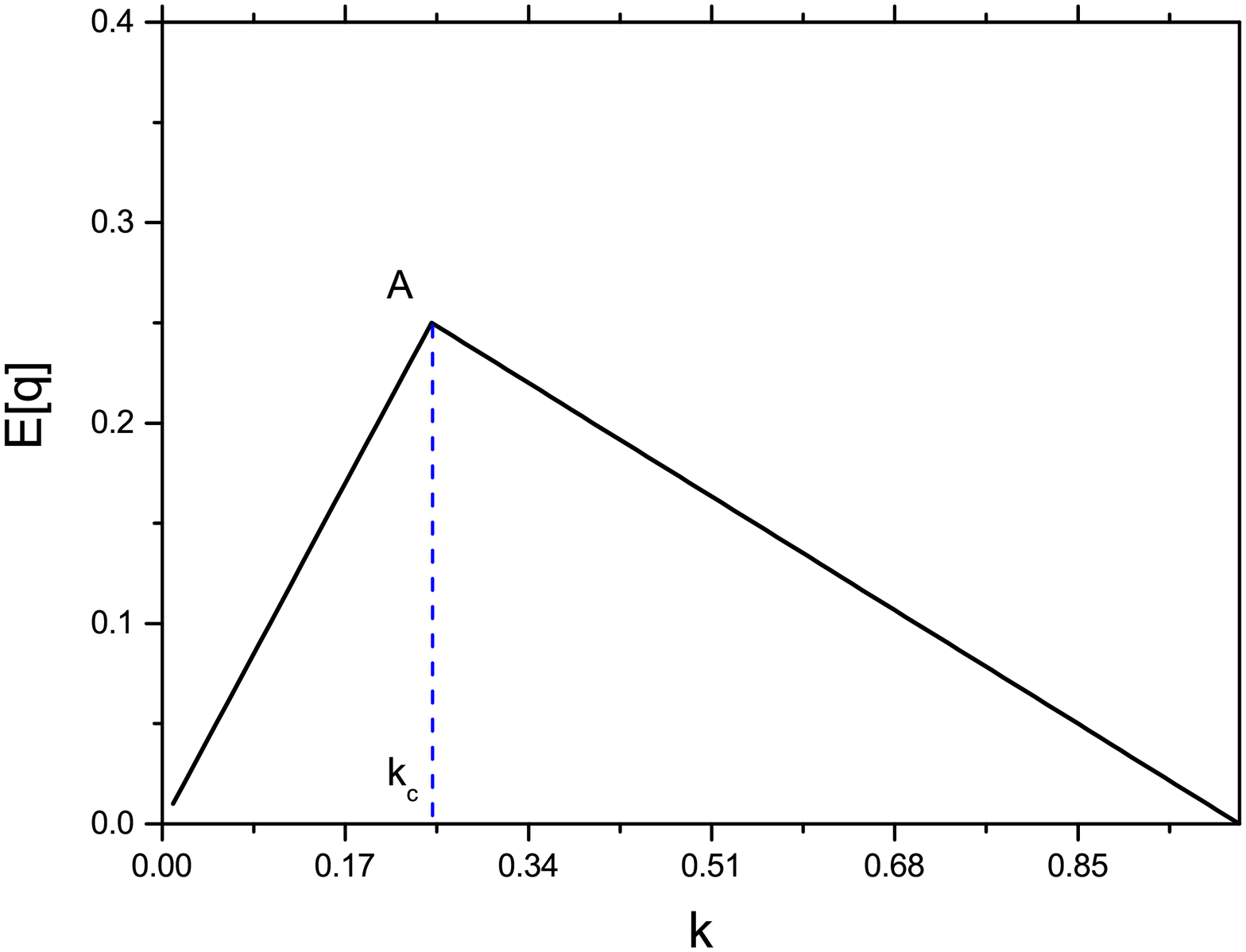} }
\end{minipage}
&
\begin{minipage}{200pt}
\centerline{\includegraphics[width=8cm]{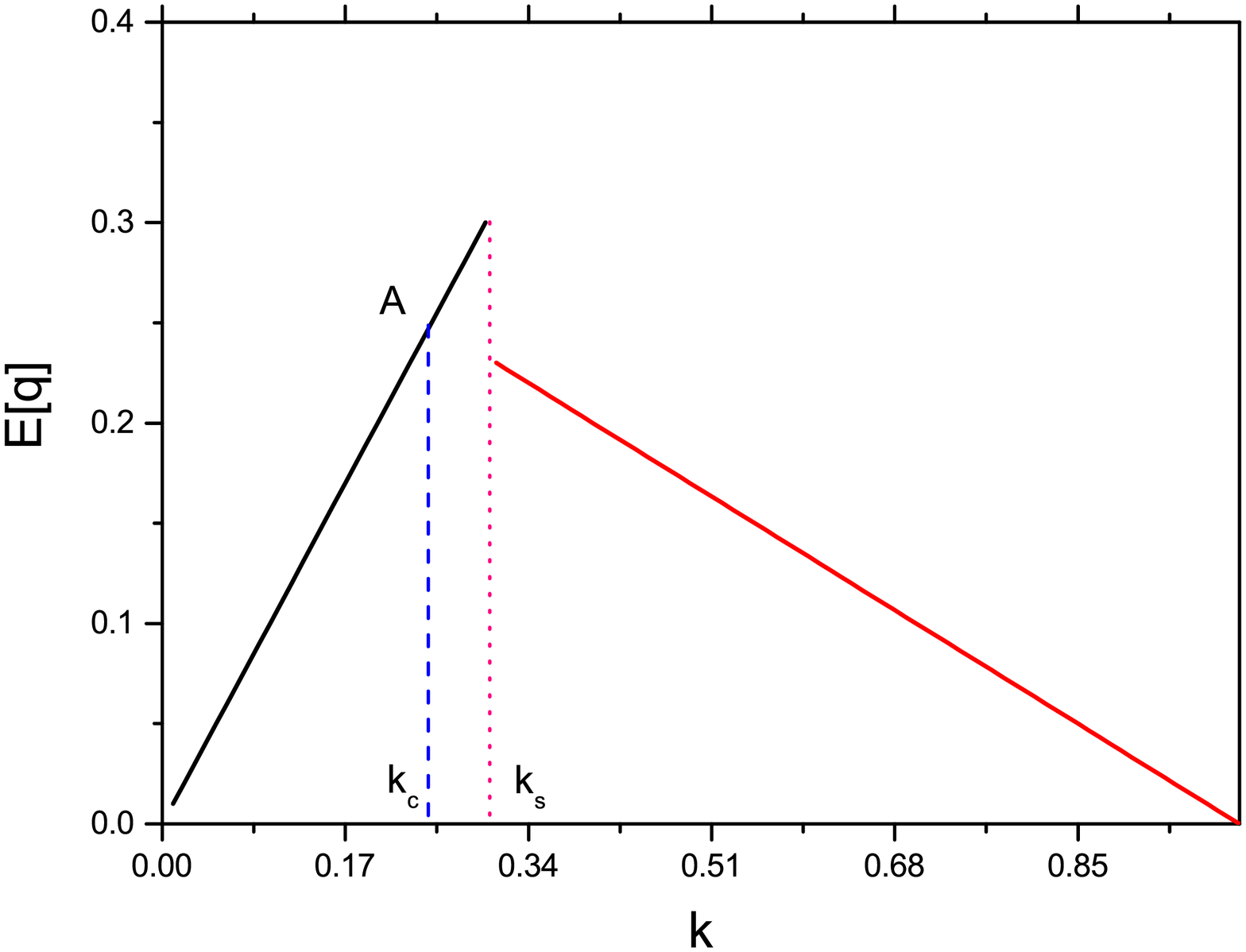}}
\end{minipage}
\end{tabular}
\caption{Schematic results from the two-speed-state model when the following trivial parametrization were adopted $v_1=0$, $v_2=c_1=N_{max}=1$, $c_2=3$.
Left: The fundamental diagram without stochastic noises. Right: The fundamental diagram when stochastic noises are considered.
The location of $k_s$ is illustrative.}
\label{fd1}
\end{figure}

The stationary solution of the model, $n^{*}$, is given by setting the r.h.s. of the above equation to zero:
\begin{eqnarray}
-c_1n_1+c_2n_1 \frac{N-n_1}{N_{max}-N}=0 ,\label{hn1}
\end{eqnarray}
and one thus obtains
\begin{eqnarray}
n^{*}_f =0 ,\label{null}
\end{eqnarray}
or
\begin{eqnarray}
 n^{*}_g=N-\frac{c_1}{c_2}(N_{max}-N) .\label{nonull}
\end{eqnarray}
One may study the stability of the stationary solution by investigating the temporal evolution via the linearized equation of small pertubations. These solutions are stable against small deviations if the stability criterion \citep{traffic-flow-btz-lob-02}
\begin{eqnarray}
 \frac{df}{dn_1}(n^{*}_{f,g})=-c_1+(c_2N-2c_2n^{*}_{f,g})\frac{1}{N_{max}-N}<0
\end{eqnarray}
is satisfied. So the null solution $n^{*}_f$ is stable only when $N< \frac{c_1}{c_1+c_2}N_{max}$, while the non-null $n^{*}_g$ is stable for $N> \frac{c_1}{c_1+c_2}N_{max}$.
Therefore, it is inferred from Eq.(\ref{q2}) (as well as from the left panel of Fig.\ref{fd1}) below that $n^{*}_f$ corresponds to the free flow solution and $n^{*}_g$ is related to the congested flow solution.
Since $\frac{c_1}{c_1+c_2}N_{max}$ plays the role of a critical density, we will introduce a parameter $N_c$ given by 
\begin{eqnarray}
 N_c=\frac{c_1}{c_1+c_2}N_{max} .\label{nc1}
\end{eqnarray}
With two speed states, the flux is given by
\begin{eqnarray}
 q=k\frac{n_1v_1+n_2v_2}{N}=\frac{1}{L}(n_1v_1+n_2v_2) ,
\end{eqnarray}
where the vehicle density $k=N/L$. 
As a result, this model predicts two distinct behaviors for the flux $q$, namely
\begin{eqnarray}
 q=\left\{ \begin{array}{ccc}
  q_{free}=kv_2  & \hspace*{2cm} & N \le N_c \\
  q_{congested}=q_c+\left[v_1-\frac{c_1}{c_2}(v_2-v_1)\right](k-k_c) & & N_c<N \le N_{max} 
 \end{array} \label{q2}
\right. ,
\end{eqnarray}
where $k_c=N_c/L$ and $q_c= k_c v_2$. The fundamental diagram in this case is a {\it continuous} curve made up of two straight lines with inclinations
$v_2$ and $v_1-\frac{c_1}{c_2}(v_2-v_1)$. 
A schematic fundamental diagram for $v_1=0$ is shown in the left panel of Fig.\ref{fd1}.
One sees that the flow arises linearly in the free flow phase $N<N_c$ and hits its peak at $N=N_c$, after which the free flow solution becomes unstable, and consequently the system transits to the second stationary solution where the flow drops linearly until it vanishes while attaining $k=k_{max}=N_{max}/L$.
The congested flow phase receives increasing contributions from the lower speed state as the vehicle density increases, while the free flow phase is solely determined by the higher speed state.
We note that the fator $\frac{1}{N_{max}-N}$ in the transition term guarantees that total congestion takes place at $k_{max}$ with $E[q]=E[v]=0$.

Now, let us briefly discuss how the free parameters of the model are related to the observed traffic characteristics.
Firstly, since the transition shall only depend on intensive quantities, the shape of the resultant fundamental diagram should be independent of $L$, the length of the highway section under consideration.
This is readily verified for the analytic solutions, Eq.(\ref{null}-\ref{q2}).
Secondly, the transition coefficients only enter the solutions for the present deterministic version of the model in terms of the ratio between $c_1$ to $c_2$.
Therefore, as far as the shape of the fundamental diagram is concerned, $c_1$ and $c_2$ can be counted as one parameter.
The speed $v_1$ corresponds to the lower-speed state, and for simplicity, it is taken to be $0$ in the present study.
Now, there are only three free paramters left, namely, $c_2$, $N_{max}$ and $v_2$.
These three parameters are used to determine the peak of the fundamental diagram (two parameters) and the maximal congestion vehicle density on the x-axis (one parameter).

Before moving on to the next section, we point out that it is not difficult to show that the above simple model is equivalent to a fold catastrophe model \citep{traffic-flow-catastrophe-book-01}, with the minima of its potential function being the roots of Eq.(\ref{hn1}).
The fold potential function is a third order polynomial:
\begin{eqnarray}
V(n_1)^{(fold)}=\frac{1}{3}\frac{1}{N_{max}-N}c_2n_1^3+\frac{1}{2}(c_1-\frac{1}{N_{max}-N}c_2N)n_1^2 ,
\label{Vfold}
\end{eqnarray}
where the stable stationary state corresponds to its minima.

\section{A stochastic fold catastrophe model}

Up to this point, we have not considered the effect of stochastic transition terms. 
The physical content of these transitions is closely related to the stochastic nature of traffic system and therefore partly provides a mathematical implementation for the scatter presented in the fundamental diagram.
Besides, as we are about to show in this section, the introduction of stochastic noises may modify the stability of the stationary solution of the fold catastrophe model.
As a result, the fundamental diagram is affected. 
It becomes discontinuous and an inverse-$\lambda$ shape appears near the maximum of the flow.
In our approach, the stochastic noises are studied by making use of SDE.
Let us introduce stochastic noises into the above fold catastrophe model as following
\begin{eqnarray}
{dn_1}&=&\left(-c_1n_1+c_2n_1 n_2\frac{1}{N_{max}-N}\right){dt}-\alpha\sqrt{c_1n_1}dB_1+\alpha\sqrt{\frac{c_2n_1 n_2}{N_{max}-N}}dB_2 ,\nonumber \\
{dn_2}&=& \left(-c_2n_2 n_1\frac{1}{N_{max}-N}+c_1n_1\right){dt}-\alpha\sqrt{\frac{c_2n_2 n_1}{N_{max}-N}}dB_2+\alpha\sqrt{c_1n_1}dB_1 ,
\label{eom1s}
\end{eqnarray}
where $B_1$ and $B_2$ are independent Brownian motions and It\^o formulas \citep{stochastic-difeq-oksendal} are assumed. \footnote{With the presence of stochastic noises, the limit of the Riemann sum does not exist any more, and in the context the It\^o formulas can be viewed as one of the two possibilities to generalize Riemann integral when the original defintions are no longer valid, and consequently a tool to solve the SDE.}
As in \citep{traffic-flow-btz-lob-01}, the stochastic transition rates are taken to be proportional to $\sqrt{n_i}$, so that the stochastic transitions weigh as much as the deterministic ones given the same vehicle accumulations \citep{stochastic-difeq-allen}.
The parameter $\alpha$ measures the strength of the stochastic noise: for $\alpha=0$, the model reduces to the deterministic counterpart. 
In this section, we restrict ourselves to the case $\alpha=1$.

\begin{figure}[!htb]
\begin{tabular}{cc}
\begin{minipage}{200pt}
\centerline{\includegraphics*[width=8cm]{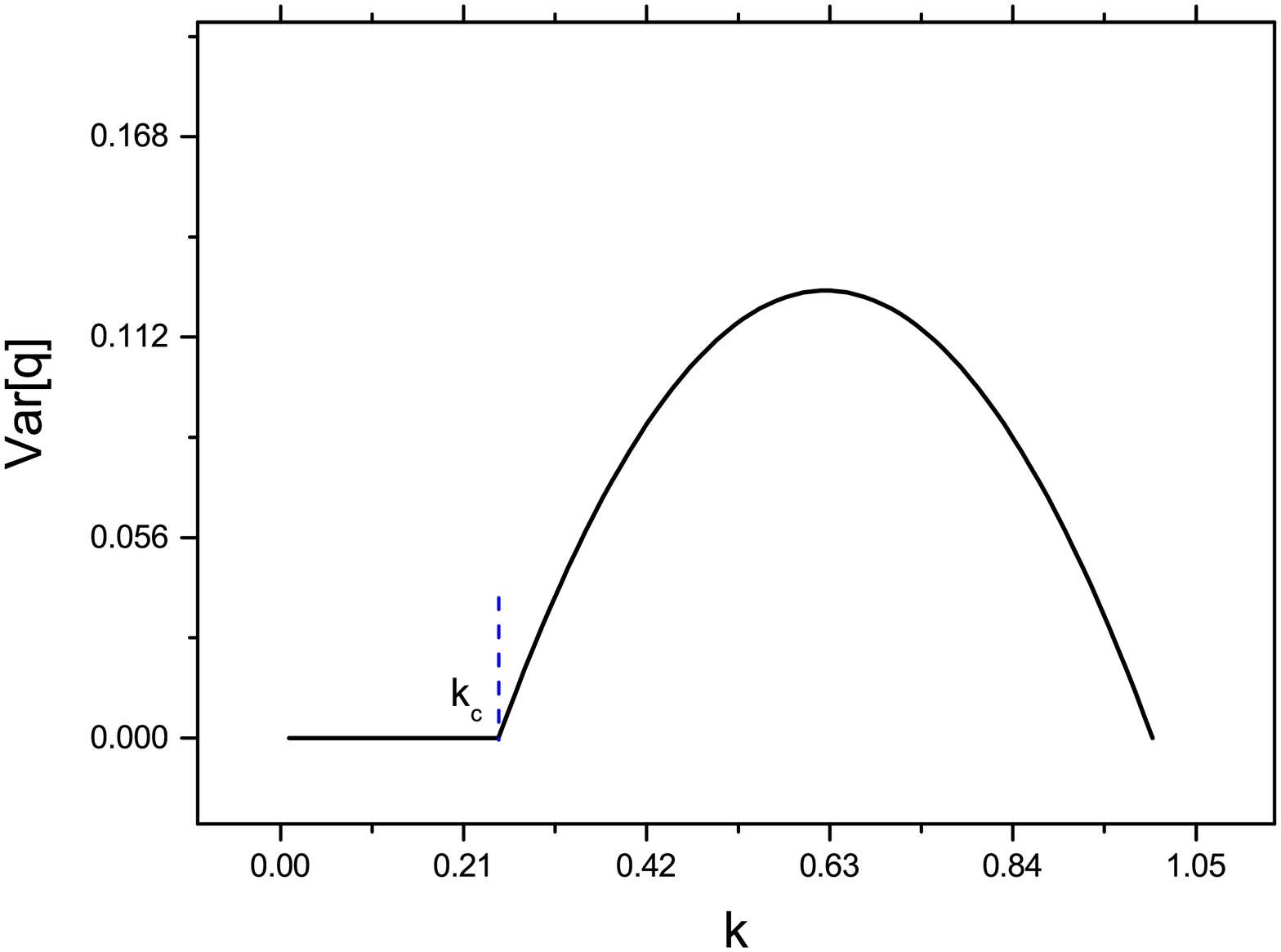} }
\end{minipage}
&
\begin{minipage}{200pt}
\centerline{\includegraphics*[width=8cm]{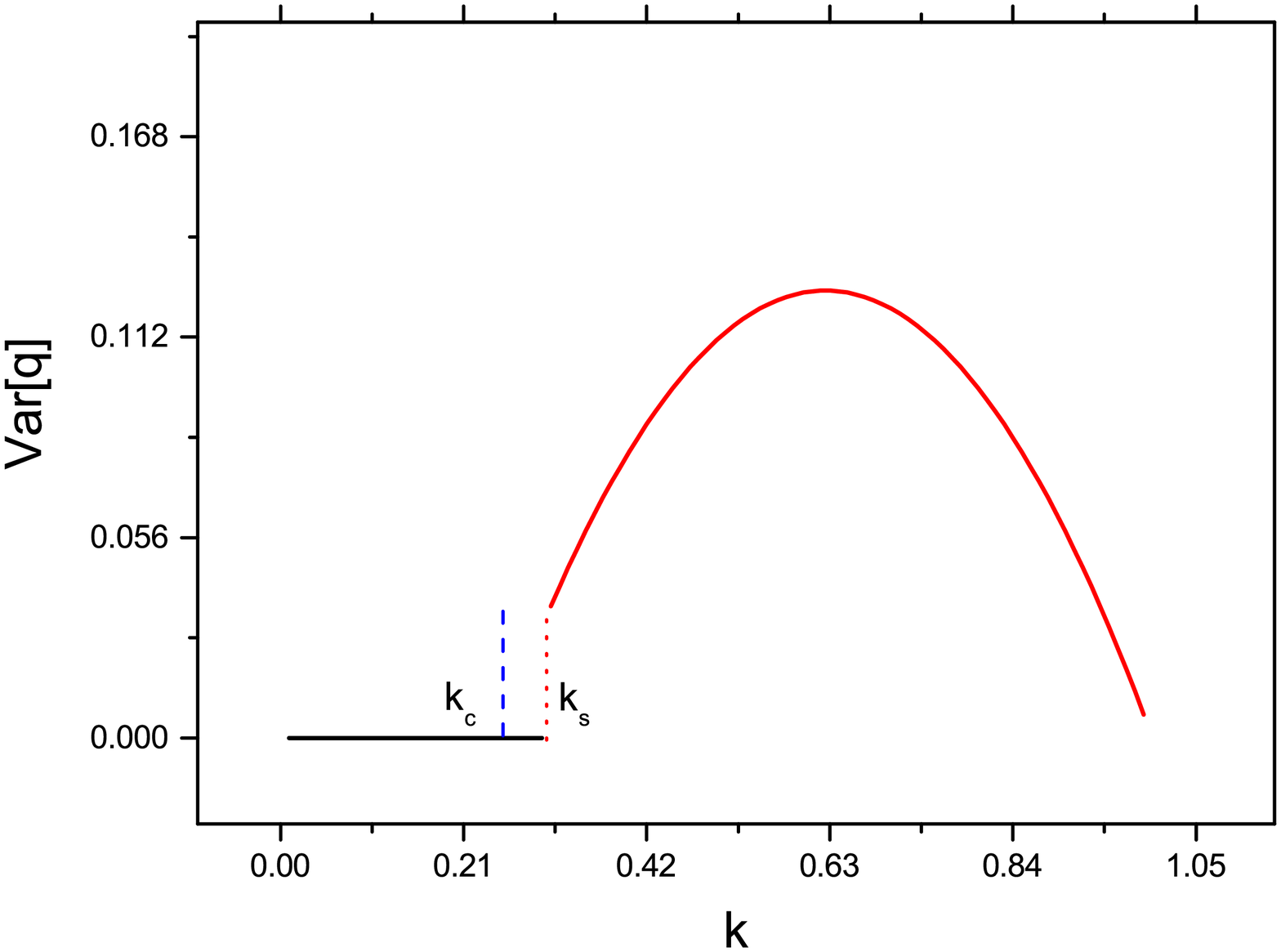} }
\end{minipage}
\\
\end{tabular}
\caption{Schematic results of the variance of the fundamental diagram of the fold catastrophe model.
In the calculation, one uses the same trivial parameters in Fig.\ref{fd1} $v_1=0$, $v_2=c_1=\alpha=N_{max}=L=1$, $c_2=3$.
In the left panel, the curves show the variances determined by Eq.(\ref{var22}); in the right panel, the stabilities of the solutions are modified, the congested flow state shrinks while free flow state extends to $k_s$ owing to the existence of stochastic noises.
The location of $k_s$ is illustrative.}
\label{varfd1}
\end{figure} 

When evaluating the expected values, it is mathematically rigorous to ignore the stochastic transition terms, as long as their coefficients (such as $\alpha\sqrt{c_1n_1}$ for instance) $\in \mathcal{V}$ (see Theorem 3.2.1 of \citep{stochastic-difeq-oksendal} for the definition of $\mathcal{V}$, which is satisfied for most well-behaved functions). 
One is then left to determine whether these stationary solutions are indeed stochastically stable. 
However, to the best of our knowledge, to obtain the analytic result on the stability of the SDE is far from a trivial task \citep{stochastic-difeq-stability-01,stochastic-difeq-stability-02,stochastic-difeq-stability-03,stochastic-difeq-stability-04}. 
In Appedix I, we analytically show that for $N<N_c$, the stable free flow solution in Eq.(\ref{q2}) for the deterministic equation, Eq.(\ref{eom1}), is indeed stable for the corresponding SDE (Eq.(\ref{eom1s})).
For other values of $N$, numerical studies are carried out to determine the stability of the stable solutions.
The results are presented in Appendix II.
It turns out that the SDE possesses different stability properties comparing to the corresponding deterministic EoM.
We found that the free flow solution is stochastically stable within the range $0 < N < N_s$, where the value $N_s$ is numerically larger than $N_c$.
When $N \ge N_s$, the system dwells around a quasi-stationary solution \citep{stochastic-difeq-grasman} for an extended period until it eventually evolves to the absorbing boundary\footnote{A type of boundary for the Fokker-Planck equation with the property that once the system evolves to it, it never comes back. It can be easily shown that $n_1=0$ is an absorbing boundary in the present problem \citep{stochastic-difeq-grasman}.}. 
Therefore, we interpret the quasi-stationary solution in the region $N_s < N < N_{max}$ in our model as the congested flow.

If one assumes that a solution is stable, one may calculate the variance of expected value following the standard procedure of It\^o calculus.
It is not difficult to show that the variances of the flow reads \citep{traffic-flow-btz-lob-02}
\begin{eqnarray}
d(n_1^2)&=&\left[-2c_1n_1^2+2c_2n_1^2 \frac{N-n_1}{N_{max}-N}\right]dt+c_1n_1dt+c_2n_1\frac{N-n_1}{N_{max}-N}dt \\
&-&2\sqrt{c_1n_1^3}dB_1+2\sqrt{c_2n_1^3\frac{ (N-n_1)}{N_{max}-N}}dB_2 ,
\end{eqnarray}
where one made use of $d(n_1^2)=2n_1dn_1+(dn_1)^2$. For steady state, one has
\begin{eqnarray}
\left[-2c_1+2c_2\frac{N}{N_{max}-N}-c_2\frac{1}{N_{max}-N}\right]{\textrm E}[ n_1^2]- 2c_2\frac{1}{N_{max}-N}{\textrm E}[n_1^3] \nonumber\\
+\left[c_1+c_2\frac{N}{N_{max}-N}\right]{\textrm E}[n_1] =0 .
\end{eqnarray}
In principle, one should further evaluate $d(n_1^3)$, which turns out to depend on higher order terms. 
For simplicity, one may employ a cut-off approximation by ignoring higher order correlation, namely, ${\textrm E}[ n_1^3] \sim {\textrm E}[ n_1] {\textrm E}[ n_1^2]$. 
This cuts the infinite equation chain at second order and one obtains
\begin{eqnarray}
{\textrm E}[n_1^2] = \frac{\left[c_1+c_2\frac{N}{N_{max}-N}\right]{\textrm E}[ n_1]}{2c_1-c_2\frac{N}{N_{max}-N}+ 2c_2\frac{1}{N_{max}-N}{\textrm E}[ n_1]} .
\end{eqnarray}
Therefore, we approximately have
\begin{eqnarray}
{\textrm {Var}}[n_1]&=& {\textrm {Var}}[n_2] = - {\textrm {Cov}}[{n_1n_2}] \nonumber \\
&=&\frac{\left[c_1+c_2\frac{N}{N_{max}-N}\right]{\textrm E}[ n_1]}{2c_1-2c_2\frac{N}{N_{max}-N}+c_2\frac{1}{N_{max}-N}+2c_2\frac{1}{N_{max}-N}{\textrm E}[ n_1]}-{\textrm E}[ n_1]^2 .
\end{eqnarray}
And the corresponding variance of the flow reads
\begin{eqnarray}
{\textrm {Var}}[q]= \frac{(v_2-v_1)^2}{L^2}\left\{\frac{\left[c_1+c_2\frac{k}{k_{max}-k}\right]{\textrm E}[ n_1]}{2c_1-2c_2\frac{k}{k_{max}-k}+c_2\frac{1}{L(k_{max}-k)}+2c_2\frac{1}{L(k_{max}-k)}{\textrm E}[ n_1]}-{\textrm E}[ n_1]^2\right\} .
\end{eqnarray}
By making use of Eq.(\ref{q2}), it can be simplified to the following form
\begin{eqnarray}
{\textrm {Var}}[q]=\left\{ \begin{array}{ccc}
  0  & \hspace*{2cm} & N \le N_c \\
  -2(v_2-v_1)^2\frac{c_1}{c_2}\left(\frac{c_1}{c_2}+1\right)(k-k_c)(k-k_{max}) & & N_c<N \le N_{max}        
 \end{array}
\right. .\label{var22}
\end{eqnarray}

In practice, the appearance of $N_s$ breaks the continuity of the fundamental diagram.
The free flow solution does not stop at $N_c$ (where it intersects with the congested flow solution at point ``A" in the left panel of Fig.\ref{fd1}) but continues until $N_s$.
Correspondingly, the congested flow solution appears as a quasi-stationary solution only when the concentrations are larger than $N_s$, which causes a gap, $\Delta q=q_{free}(N_s)-q_{congested}(N_s)$, at $N_s$, as can be inferred from Eq.(\ref{q2}). This is shown in the right panel of Fig.\ref{fd1}, in comparison to the deterministic case shown in the left panel of the same figure.

In Fig.\ref{varfd1}, we present the schematic results of the variation of the fundamental diagram by using the same parameters as those in Fig.\ref{fd1}. 
It is shown that the variance is zero in the free flow phase because ${\textrm E}[ n_1]=0$. 
Then the variance increases when the system enters the congested phase (as an unstable solution at first when $N_c < N < N_s$).
It reaches the maximum and eventually decreases to zero when the complete congestion occurs, as seen from Eq.(\ref{var22}).
However, we note that the calculated variance is not applicable to the transition region $N\sim N_s$, where traffic states are not stable.
This point will be further discussed below.
We emphasize that the obtained capacity drop is not a consequence of any additional and/or artificial parameter of the model, but from the modified stability owing to the existence of stochastic noises. 

\section{Qualitative comparison to the data}

In this section, we carry out a qualitative comparison to the I-80 data from the NGSIM program \citep{traffic-flow-data-08} for model validation.

The public I-80 data is collected from the six-lane freeway section on I-80 in the San Francisco Bay area in Emeryville, CA, with an on-ramp from Powell Street located within the study area. 
It includes three intervals of 4 pm to 4:15 pm, 5 pm to 5:15 pm and 5:15 pm to 5:30 pm on April 13, 2005. 
The original data set consists of video clips captured by seven synchronized digital video cameras, which are then transcripted into vehicle trajectories.
The latter provide the precise locations of all vehicles within the study area every one-tenth of a second, for a total of 5408 different automobiles.
The present analysis only concerns automobiles (neither motorcycles nor trucks), which corresponds to roughly 96\% of all vehicles.
Since lane 1 (the leftmost lane of the freeway) is a high-occupancy vehicle (HOV) lane, its inflow is restricted especially during the observation period and consequently behaves completely different than the other five lanes.
For this reason, the data from lane 1 is excluded.
We also note that the original study by Cassidy \citep{traffic-flow-data-inverse-lambda-05} used loop detector data, which are obtained between 500m - 2000m downstream of an on-ramp where a bottleneck is formed during the relevant period.
To mimic the loop detector data, our calculations are carried out for fixed spatial positions downstream of the Powell Street on-ramp.

It is noted that conventional measures in the data aggregation process, for instance, the time mean speed and space mean speed, usually present several sources of noise.
The latter impose considerable difficulty in our current study, due to the lack of statistics.
In this respect, we employ the method proposed by Coifman \citep{traffic-flow-data-18} over the conventional one.
This is because Coifman's method was shown to be able to efficiently suppress the noises, such as unphysical fluctuations related to the integer nature of vehicle number.
Tailored to our present study, moreover, an additional criterion is proposed to identify the near-stationary states.

\begin{figure}[!htb]
\begin{tabular}{cc}
\begin{minipage}{200pt}
\centerline{\includegraphics*[width=8cm]{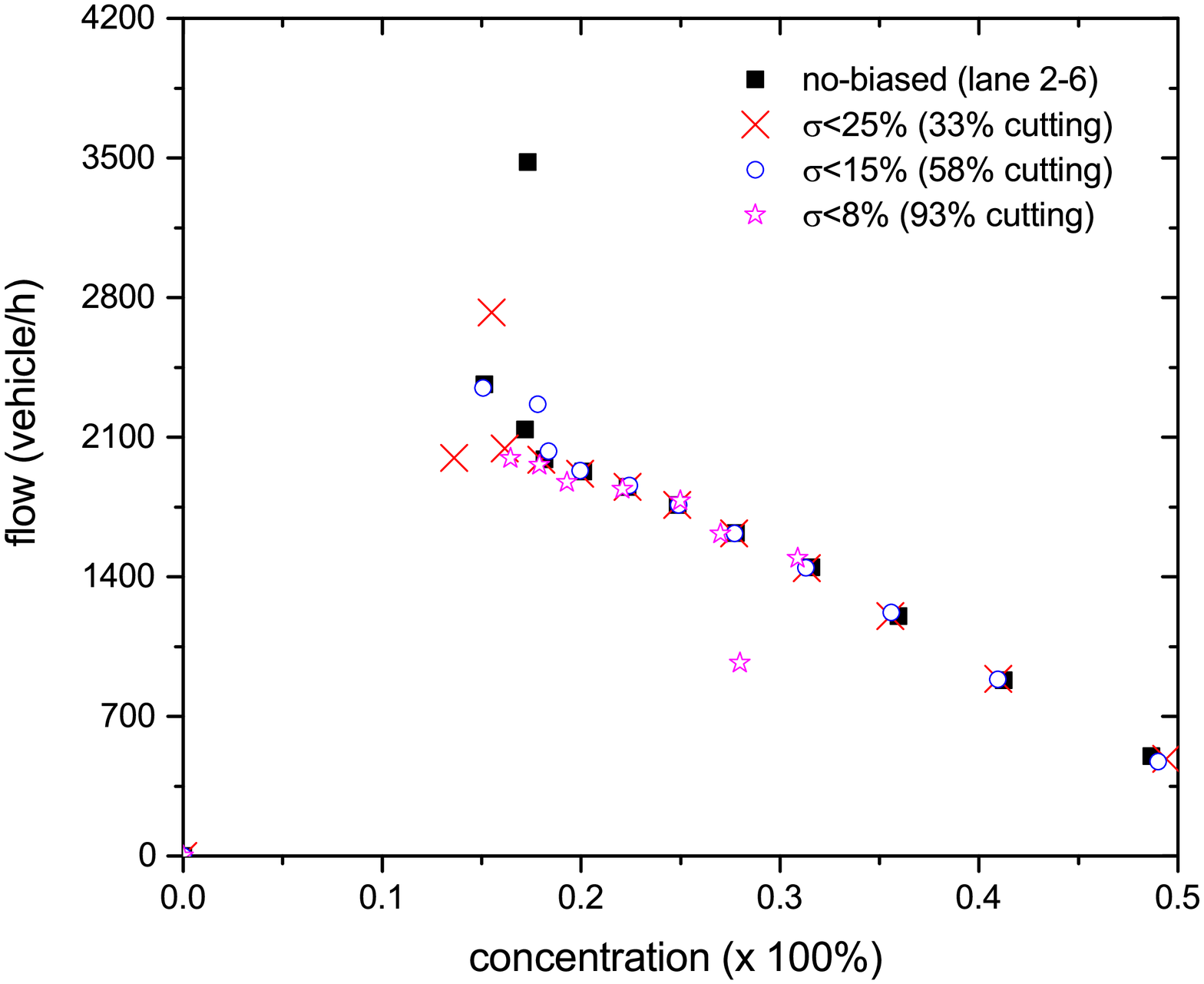} }
\end{minipage}
&
\begin{minipage}{200pt}
\centerline{\includegraphics*[width=8cm]{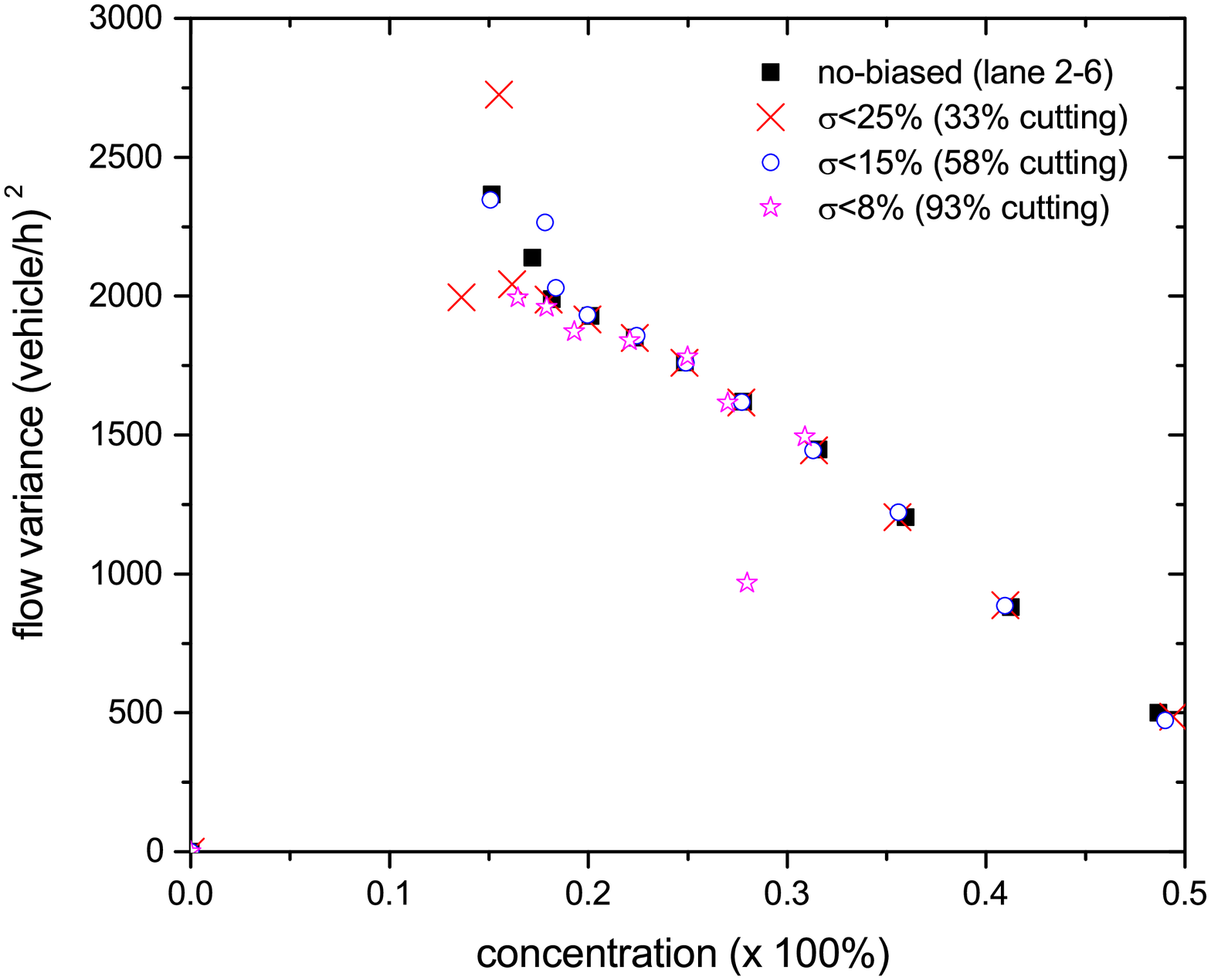} }
\end{minipage}
\end{tabular}
\caption{Left panel: fundamental diagram calculated by using the method in \citep{traffic-flow-data-18} with different value of $sigma$ introduced in the text; right panel: the corresponding flow variance. The data from lane 1 is excluded from the calculations.}
\label{i80fit-05}
\end{figure}

According to Coifman's method, the fundamental diagram is evaluated as follows.
Primarily, the traffic state is measured in terms of the headway for each vehicle.
The resulting data points are then grouped by similar vehicle lengths and speeds before aggregation.
In particular, individual vehicle headway $h$ and vehicle length $l$ are used to calculate the concentration, $k=l/h$, during single vehicle passage.
The corresponding flow is evaluated by $q=h/v$ where $v$ is the measured speed of the vehicle. 
The above-calculated flow-density pair $(q, k)$ gives rise to an individual data point.
To further improve the statistics, we consider 50 distinct spatial points including all the data starting at the immediate downstream of the on-ramp from Powell Street. 
A tiny portion of data give unreasonably small headway (less than the vehicle length) and therefore are not considered in our calculations. 
Next, the data points are sorted into smaller speed bins.
By calculating the average quantities in each bin, the flow concentration relation can be obtained, as shown in filled black squares in the left panel of Fig.\ref{i80fit-05}.
It is found that the latter present a reasonably good resolution, especially for the congestion phase. 

Although Coifman's original method does not involve the concept of the near-stationary condition, we now explicitly add this ingredient into the data analysis by introducing a criterion as follows.
An individual data point (the flow-density pair $(q, k)$ defined above) will be taken into account, only when the speed variance of successive vehicles is small enough within a given time interval. 
Here the time interval under consideration is centered by the instance of the vehicle passage in question.
To be more specific, we define,
\begin{eqnarray}
\sigma = \frac{\sqrt{Var[v]}}{E[v]}
\end{eqnarray}
to be the ratio between the standard deviation to the average speed, and use it as a measure for the threshhold of the near-stationary states.
The time interval is taken to be of two minutes in our analysis.
The resulting fundamental diagrams with different values of $\sigma$ are shown, additionally, in the left panel of Fig.\ref{i80fit-05} in various symbols (namely, red x marks, empty blue circles, and empty purple stars respectively).
In the right panel of Fig.\ref{i80fit-05}, the corresponding flow variances are presented.

One finds that for unbiased data shown in filled black squares in the left panel of Fig.\ref{i80fit-05}, the maximum of the free flow phase arises above the congestion phase, forming an inverse-$\lambda$ shape, which is consistent with previous data analysis and the stochastic fold catastrophe model.
As the $\sigma$-cut becomes more stringent, the selected flow states become closer to the near-stationary ones. 
Meanwhile, the inverse-$\lambda$ shape starts to disappear as it approaches the deterministic fold catastrophe model.
This confirms the findings in \citep{traffic-flow-data-inverse-lambda-05}.
However, a severe $\sigma$-cut may remove a significant portion of the data points, which in turn decreases the data resolution.
The right panel of Fig.\ref{i80fit-05} shows that the variance of the flow diminishes with increasing vehicle concentration. 
This is intuitive as the vehicle speed vanishes for a complete congestion.

Now, we are in a position to calibrate the model parameters to reproduce the obtained flow concentration relation.
Similar to the deterministic case discussed above, the model calibration is implemented by fitting $v_2$ by the inclination of the free flow state, the ratio between $c_1$ to $c_2$ by the maximum flow of the stable state, and $N_{max}$ by the congestion vehicle density.
In other words, the three free parameters of the model are determined by the essential characteristics of a fundamental diagram, namely, the free flow speed, the maximum of the flow and the maximal congestion vehicle density.
This implies that similar calibration procedures can be easily carried out for different traffic scenarios, as long as the concept of the fundamental diagram is applicable.

\begin{figure}
\begin{tabular}{cc}
\begin{minipage}{200pt}
\centerline{\includegraphics*[width=8cm]{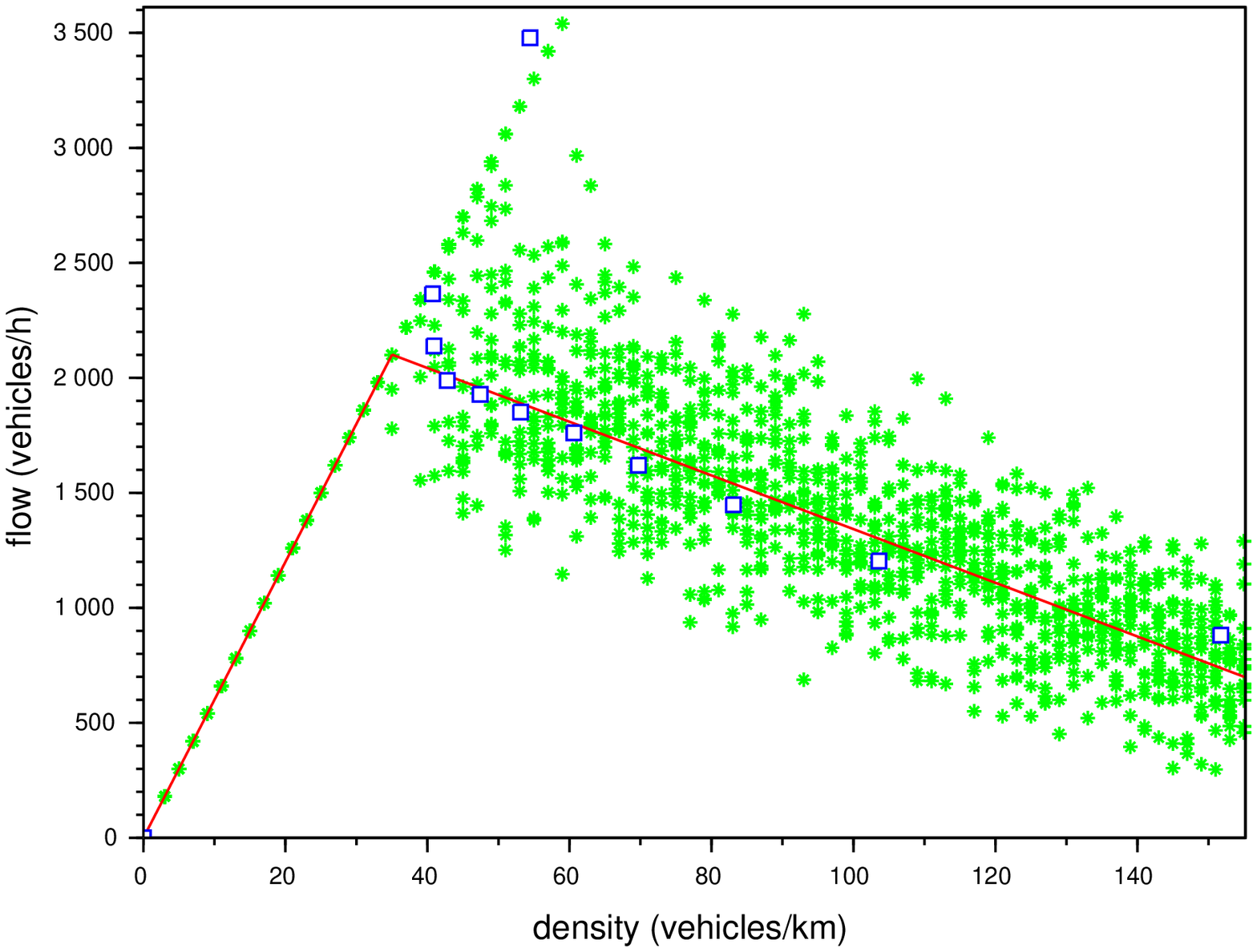} }
\end{minipage}
&
\begin{minipage}{200pt}
\centerline{\includegraphics*[width=8cm]{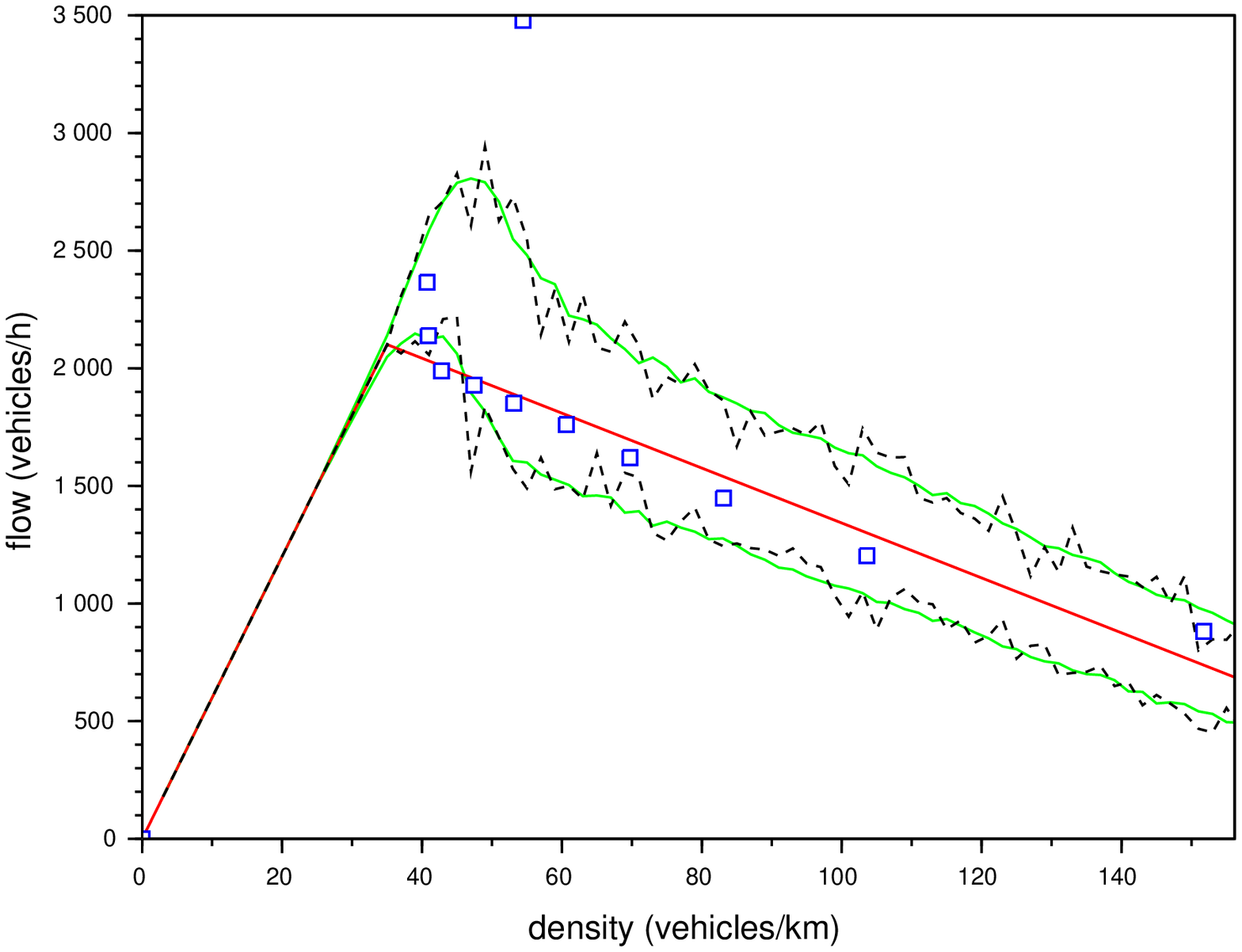} }
\end{minipage}
\end{tabular}
\caption{The result of model calibration. 
The fundamental diagram obtained by the using $c_1=1, c_2=5.14, L=1 km, N_{max}=215, v_1=0, v_2=60 km/h$.
The solid red lines are the average flow by Eq.(\ref{q2}) while the empty blue squares present the data shown in Fig.\ref{i80fit-05}.
Left: For each of the 120 given values of $N$ (thus the vehicle concentration), 20 simulations are carried out with initial condition $n_1(t=0)=N/8$.
Each filled green star corresponds to 1 simulation, which is taken (measured) at a sufficiently long time $t=20$ with $\Delta t=0.01$. 
Right: The standard deviations are calculated numerically for each given $N$ by using 20 and 1000 simulations, indicated by dotted black curves and solid green curves respectively.
We note that Coifman's method sort the data points into speed bins, two different speed bins may have the same value of concentration, which in turn affects the calculated average flow.}
\label{i80fit-06}
\end{figure}

The results are shown in Fig.\ref{i80fit-06}, where the model parameters are calibrated by using Fig.\ref{i80fit-05}.
Each simulation is represented by a filled green star in the left panel of Fig.\ref{i80fit-06} and the data from Fig.\ref{i80fit-05} are shown in empty blue squares.
It is found that for the free flow states, different simulations coincide with one another, and the resultant variance is minimal.
This is in accordance with our analytic results in Eq.(\ref{q2}) and Eq.(\ref{var22}).
The free flow phase stretches beyond the transition vehicle density and forms an inverse-$\lambda$ shape with the congestion states.
The congested flow phase, on the other hand, displays significant complication owing to the reasons discussed previously.
In the region with high concentration, the average flow, as well as its standard deviation, are described by Eq.(\ref{q2}) and Eq.(\ref{var22}).
As observed in the right panel of Fig.\ref{i80fit-05}, they both decrease as the vehicle density increases and approaches that of complete congestion.
In the transition region, there is an interplay between the free flow phase and congested flow phase with stochastic uncertainties.
Fig.\ref{i80fit-05} shows that for a given value of vehicle density, the system may end up being observed on the free flow branch or the congestion flow branch.
And the bifurcation, in turn, affects the average flow and gives rise to a much larger deviation when the data were taken indiscriminately.
In this context, the average flow might not be a good measure for the particular data analysis employed in this study.
As a matter of fact, Coifman's method sort the data points into speed bins, two different speed bins may have the same value of concentration, as observed from Fig.\ref{i80fit-06}.
We show the calculated standard deviation by using different numbers of simulations in the right panel of Fig.\ref{i80fit-06}.
The calculated standard deviation of the flow is finite and it converges as the number of simulations increases.
Overall, our numerical calculations are in qualitative agreement with the data, the observed complexity in the transition region is reasonably reproduced.

\section{Discussions and concluding remarks}

In the previous section, it is shown that the present model can qualitatively reproduce the observed scatter in the data.
We understand that the scatter presented in the numerical results is due to the transition between the two phases, namely, the free flow and congested flow, in addition to the flow variance of the congested flow.
Although these two phases are both locally stable in the deterministic version of the model, the presence of the stochastic noises modifies the stability and triggers the transitions among them.
However, it is worth noting that the scatter around the transition region may also be caused by other mechanisms besides stochastic noises.
For instance, the limit cycle solution forms a closed trajectory in phase space; it may also lead to the scatter in the data even in a purely deterministic system \citep{traffic-flow-btz-lob-02}.
In practice, in order to differentiate a deterministic temporal evolution from stochastic noises, we have to pay specific attention to the different time scales of the two phenomena.
White stochastic noises, by definition, appear consistently. 
If they are suppressed for some reason, one should be able to observe a near-stationary state which slowly evolves in time.
In other words, a deterministic phenomenon usually plays a more significant role at larger time scale than stochastic fluctuations.
Therefore, these two different phenomena can be separated by a careful analysis of the variance of physical observables at various time scales.
Numerical simulation provides another possibility since in this case, one can manually interfere by turning off certain sources.
In this context, the effect of individual mechanism can be investigated with or without the presence of others.
For this very reason, at the current stage of the investigation, we do not attempt to reproduce the data quantitatively as the present model is not yet complete.
Our finding is that the effect of stochastic noises on the stability of the EOM provides one of the possibilities.

We also point out that the present approach does not contradict the three-phase traffic theory \citep{traffic-flow-three-phase-01,traffic-flow-three-phase-02}.
In the latter case, the scattered data on a flow-density plot can be achieved by introducing a two-dimensional meta-stable region.
Concerning the present model, the properties of the so-called {\it synchronized flow} are reasonably captured by the quasi-stationary solution.
This is because a quasi-stationary solution of stochastic differential equation stays in the neighborhood of the equilibrium point for an extended period.
It maintains its time mean properties and occupies a two-dimensional region in the fundamental diagram due to its temporal oscillations (see the bottom right plot in Fig.\ref{nstability} in Appendix II).
Therefore, the quasi-stationary state located below the free flow state can be identified as the synchronized flow (e.g., \citep{traffic-flow-data-04}).

To summarize, in this work we investigate a nonlinear mesoscopic model for the fundamental diagram of traffic flow.
Regarding the potential function, our approach is equivalent to the well-known catastrophe model.
We found that the inverse-$\lambda$ shape and the associated sudden jump in the fundamental diagram can be attributed to the modified stability of the system triggered by stochastic noises. 
As discussed in this paper, the latter causes the co-existence of different stable and/or near-stationary traffic flow states, which subsequently leads to the observed feature in the fundamental diagram.
In particular, with the presence of stochastic noises, the free flow state stretches beyond the maximum point of its deterministic counterpart. 
Meanwhile, the congested flow phase shrinks toward the region of high concentration.
As a result, a vertical gap appears below the maximum of the flow, and an inverse-$\lambda$ shape is consequently formed.
Interestingly, the above inverse-$\lambda$ shape shall not appear at all if one only considers stationary traffic states.
This is because a stationary traffic state is equivalent to the situation where the strength of the random fluctuations is largely suppressed, in other words, the model restores to its deterministic version.
In this context, the capacity drop in the fundamental diagram is dynamical.
By model calibration, we show that the present approach, with a minimal number of parameters, captures the main characteristics of the data.

\section{Acknowledgements}

We acknowledge the financial support from Funda\c{c}\~ao de Amparo \`a Pesquisa do Estado de S\~ao Paulo 
(FAPESP), Funda\c{c}\~ao de Amparo \`a Pesquisa do Estado de Minas Gerais (FAPEMIG),
Conselho Nacional de Desenvolvimento Cient\'{\i}fico e Tecnol\'ogico (CNPq),
and Coordena\c{c}\~ao de Aperfei\c{c}oamento de Pessoal de N\'ivel Superior (CAPES).

\section{Appendix I: stability of the SDE for $N<N_c$}

Here we show that the free flow solution $x(0,t) \equiv n_1(0,t)=0$ of the Eq.(\ref{eom1}) is {\it stochastically stable} (Definition 2.1) by using Theorem 2.2 of Chapter 4 of the reference \citep{stochastic-difeq-xmao}.
The outline of the proof is as follows. 
To show that the solution $x(t)\equiv 0$ is stable for a stochastic equation
\begin{eqnarray}
dx(t) = f(x(t),t)dt +g(x(t),t)dB \nonumber
\end{eqnarray}
one needs to find a positive-definite function $V(x,t) \in C^{2,1} (S_h \times [t_0, \infty);R_+)$ such that $LV(x,t) \le 0$.
Here $V(x,t)$ corresponds to the Lyapunov function of an ordinary differential equation, which measures the {\it distance} of a small perturbation from the equilibrium solution; and
\begin{eqnarray}
L=\frac{\partial }{\partial t}+\sum_i f_i(x,t)\frac{\partial}{\partial x_i}+\frac{1}{2}\sum_{i,j}[g(x,t)g^T(x,t)]_{i,j}\frac{\partial^2}{\partial x_i\partial x_j} \nonumber
\end{eqnarray}
is a differential operator.
$LV(x,t)$ bears the interpretation of time derivative of the distance $V$. Besides, for the theorem to be valid, one needs $f(0,t)=g(0,t)=0$, which is readily satisfied by the equation of $n_1$
\begin{eqnarray}
{dn_1}&=&\left(-c_1n_1+c_2n_1 (N-n_1)\frac{1}{N_{max}-N}\right){dt}-\sqrt{c_1n_1}dB_1+\sqrt{\frac{c_2n_1 (N-n_1)}{N_{max}-N}}dB_2 \label{eom1_ff}
\end{eqnarray}
Now let us take $V(x,t)=x$. It is straightforward to show that $V$ is indeed positive definite and
\begin{eqnarray}
LV(x,t) = -c_1x+c_2x (N-x)\frac{1}{N_{max}-N} \nonumber
\end{eqnarray}
It is straightforward to show that the condition $LV \le 0$ implies $N< N_c$, with $N_c$ defined in Eq.(\ref{nc1}).

\section{Appendix II: numerical study of the stability of the SDE}

In this Appendix, we numerically investigate the stability of the expected value as well as the variance of the free flow and congested flow solutions of Eq.(\ref{eom1}) according to their definitions. 
The equation for free flow is Eq.(\ref{eom1_ff}). One may also write down the equation for the congested flow
\begin{eqnarray}
{d\tilde{n}_1}=\left(c_1\tilde{n}_1-c_2\tilde{n}_1 (N+\tilde{n}_1)\frac{1}{N_{max}-N}\right){dt}-\sqrt{c_1(\tilde{n}_1+n^*_g)}dB_1+\sqrt{\frac{c_2(\tilde{n}_1+n^*_g)(N-(\tilde{n}_1-n^*_g))}{N_{max}-N}}dB_2 \nonumber
\end{eqnarray}
where one makes use of $n_1 \rightarrow {\tilde {n}}_1=n_1-{n^{*}}_g$, such that $\tilde{n}_1=0$ corresponds to the congested flow solution and the stochastic stablity can be verified according to its definition.
We note that the above equation is not symmetric in comparison to Eq.(\ref{eom1_ff}), since the former has $\tilde{f}(0,t)=0$ and $\tilde{g}(0,t) \ne 0$. 
The parameter space of the system is one-dimensional. 
Therefore it is sufficient to present the numerical results of the evolution of free flow as shown in Fig.\ref{nstability} below.

\begin{figure}[!htb]
\begin{tabular}{cc}
\begin{minipage}{200pt}
\centerline{\includegraphics*[width=8cm]{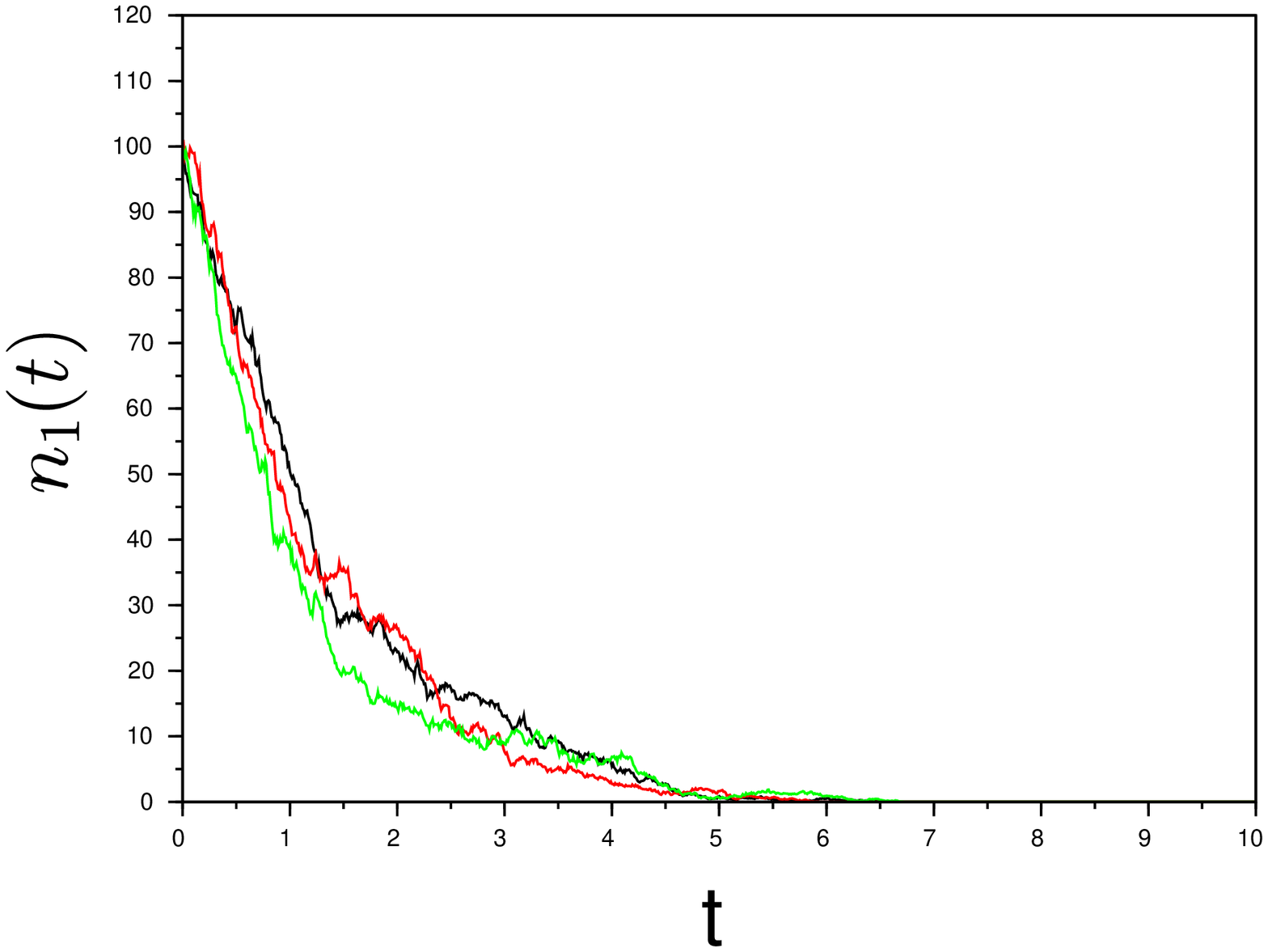}}
\end{minipage}
&
\begin{minipage}{200pt}
\centerline{\includegraphics*[width=8cm]{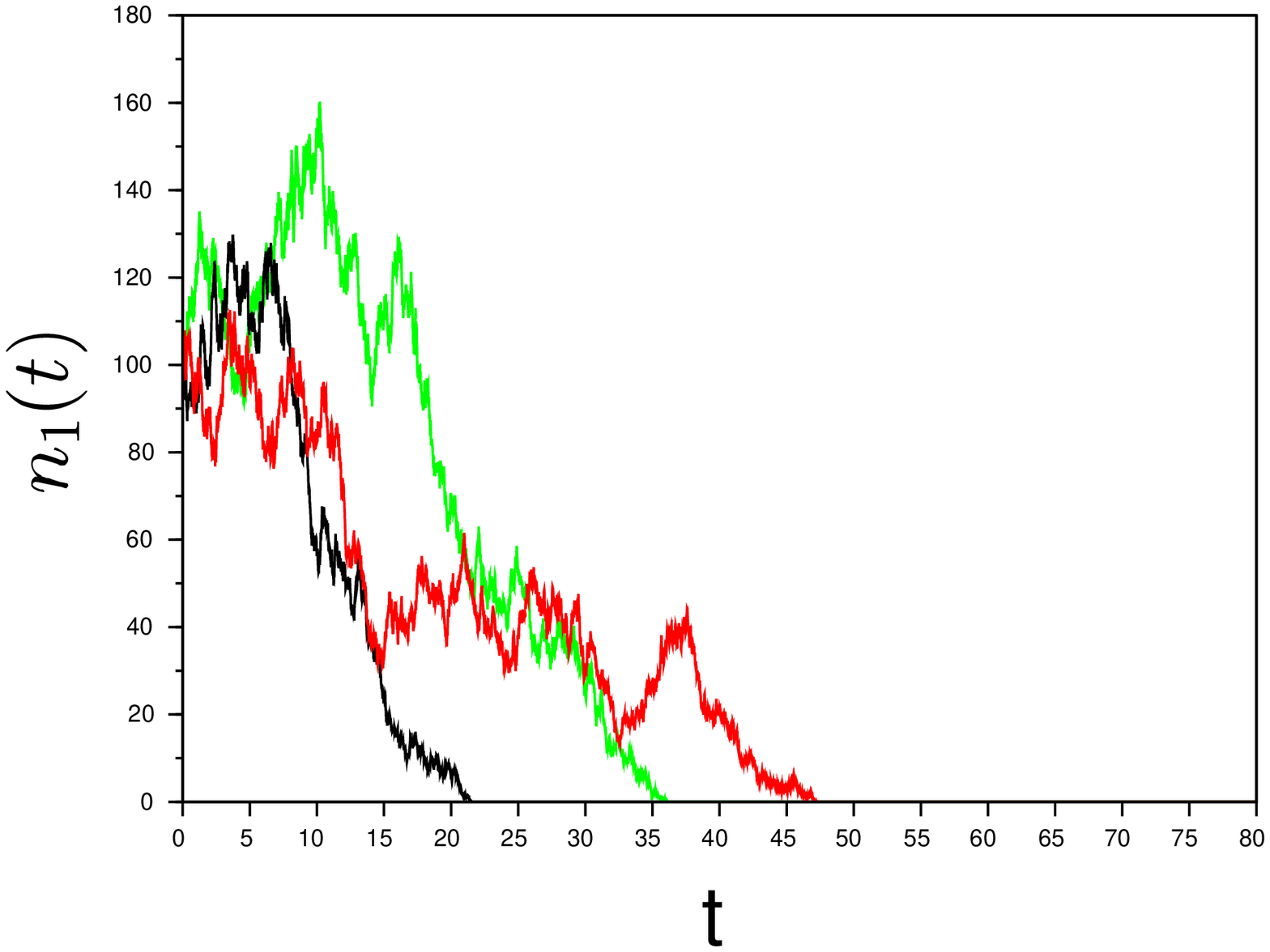}}
\end{minipage}
\\
\begin{minipage}{200pt}
\centerline{\includegraphics*[width=8cm]{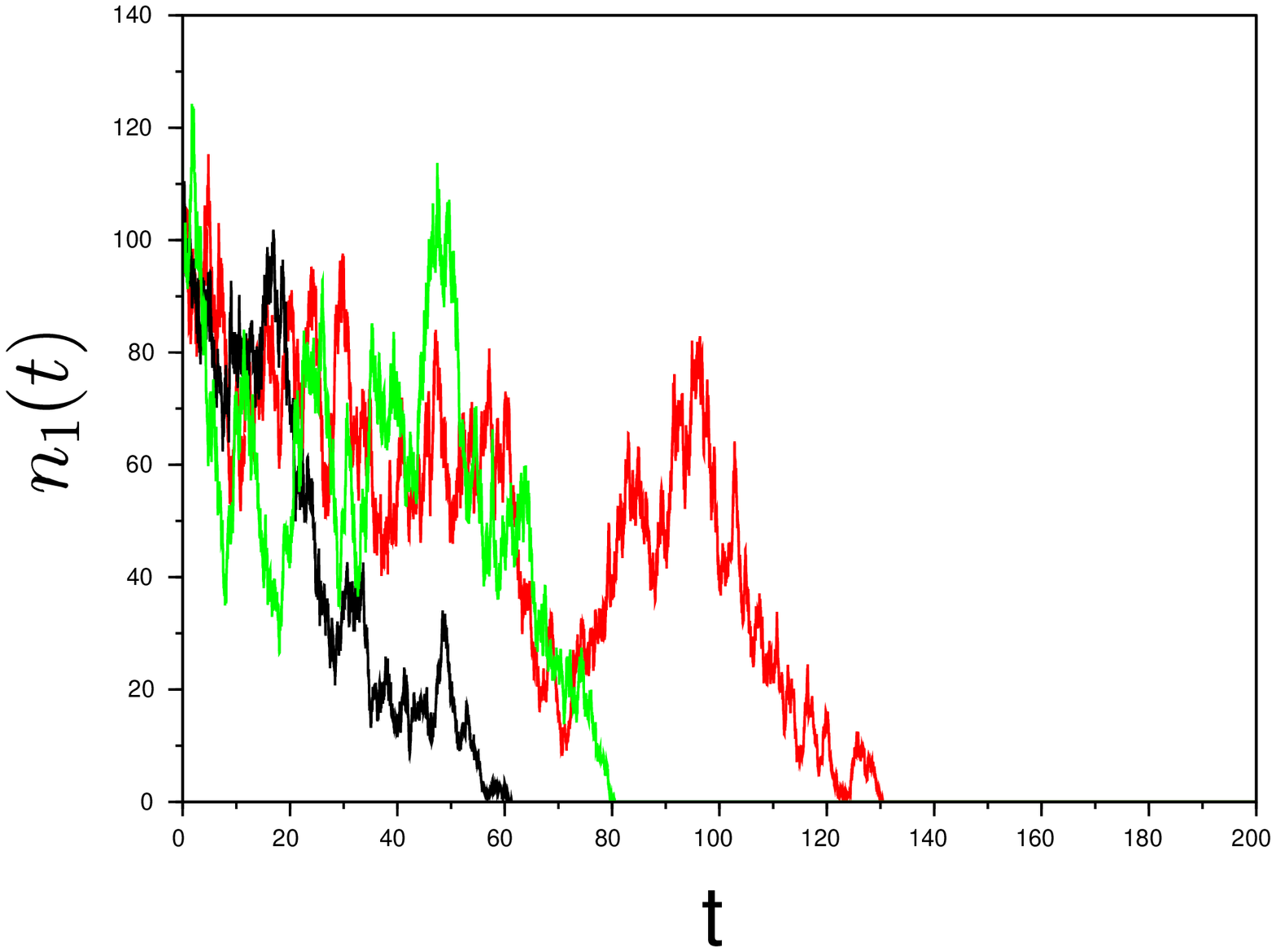}}
\end{minipage}
&
\begin{minipage}{200pt}
\centerline{\includegraphics*[width=8cm]{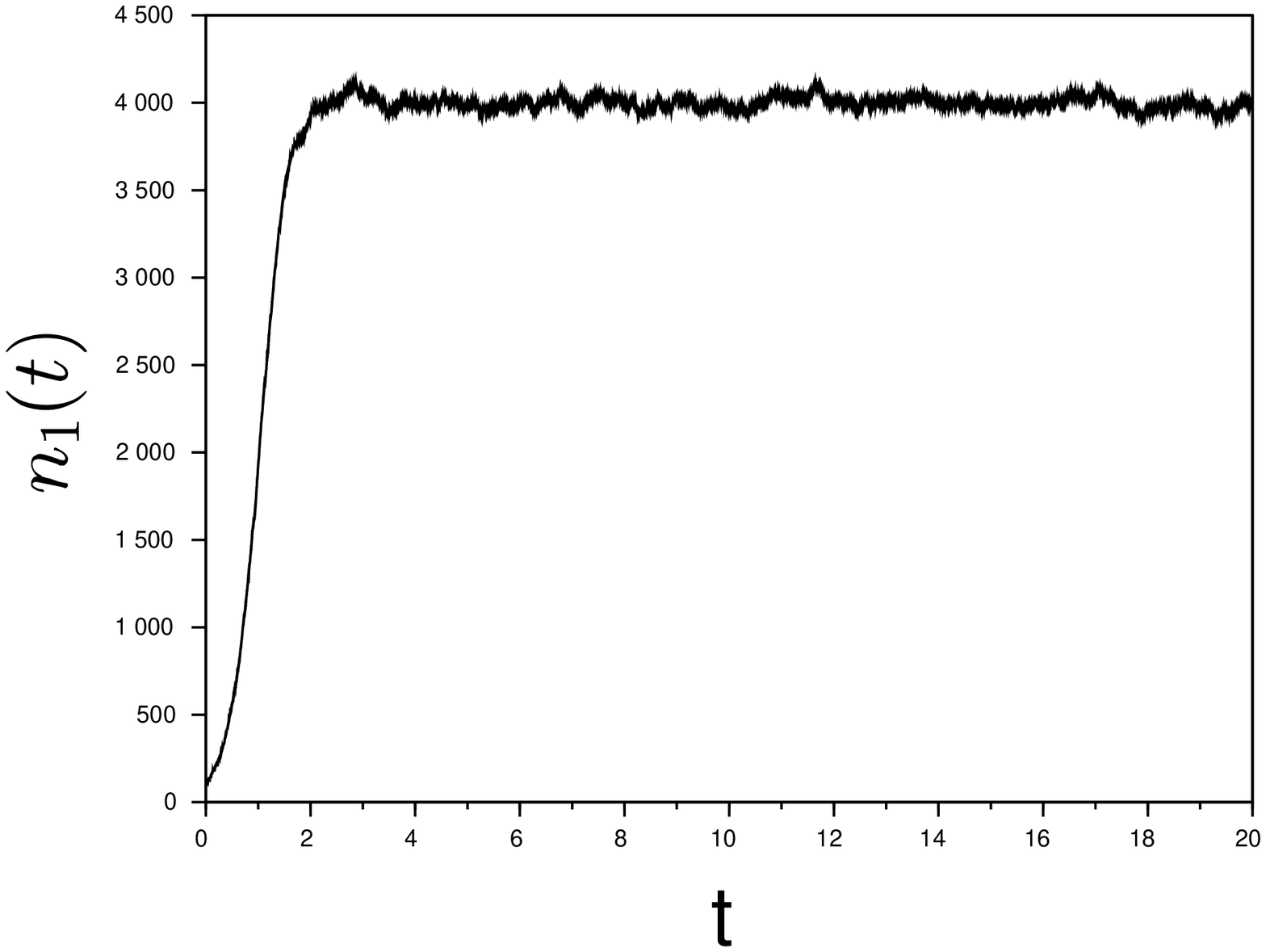}}
\end{minipage}
\end{tabular}
\caption{Numerical study of the stability of the SDE for the free flow and the congested flow. The calculations are carried out using the same parameters of the fit shown in Fig.\ref{i80fit-05}, one has $N_c=2204$.
The plots show the temporal evolution of small perturbations ($n_1(0)=100$) for different road concentrations $k=N/L$. 
Each plot illustrates (in different colors) several of many numerical simulations carried out.
Top left: temporal evolution for small perturbations for the free flow state with $N=1000<N_c$; 
Top right: the same as the top left plot but with $N=N_c$;
Bottom left: the same as the top left plot but with $N_c<N = 2214 <N_s = 2360$, all the three plots show that numerically $n_1(t \rightarrow +\infty)= 0$, therefore the solution is stochastically stable as long as $N < N_s$;
Bottom right: the same as the top left plot but with $N = 5204 >N_s$, it shows that the system evolves to a quasi-stationary state as $n_1 \rightarrow n^*_g$.}
\label{nstability}
\end{figure} 

Fig.\ref{nstability} shows that both the expected value and variance of the free flow is stochastically stable (approaches zero) when $N<N_c$ as proven in Appendix I.
However, the range of stable free flow actually reaches beyond $N_c$. In fact, free flow is stable when $N<N_s$ with $N_s > N_c$.
When $N>N_s$, the system evolves towards the equilibrium $n_1=n^*_g$ (Eq.(\ref{nonull})) and stays in its neighborhood for a very long period before it eventually evolves to the absorbing boundary.
It is noted that such temporal evolution is stochastic in nature and therefore not periodic, it is known as a quasi-stationary solution according to \citep{stochastic-difeq-grasman}.
Since $n^*_g$ corresponds to the congested flow state in the deterministic fold catastrophe model, we, therefore, identify this solution as the congested flow state.
Numerically one finds $k_s = k_c + 15.5 \pm 0.5$.

\bibliographystyle{ormsv080}

\bibliography{references_qian}{}

\end{document}